# Clocking plasmon nanofocusing by THz near-field streaking

Lara Wimmer[1], Benjamin Schröder[1], Murat Sivis[1], Georg Herink[1], and Claus Ropers[1]

IV. Physical Institute – Solids and Nanostructures, University of Göttingen, Friedrich-Hund-Platz 1, 37077 Göttingen, Germany

Correspondence to: claus.ropers@uni-goettingen.de

**We apply terahertz (THz) near-field streaking in a nanofocusing geometry to investigate plasmon polariton propagation on the shaft of a conical nanotip. By evaluating the delay between a streaking spectrogram for plasmon-induced photoemission with a measurement for direct apex excitation, we obtain an average plasmon group velocity, which is in agreement with numerical simulations. Combining plasmon-induced photoemission with THz near-field streaking facilitates extensive control over localized photoelectron sources for time-resolved imaging and diffraction.**

Propagating surface plasmon polaritons (SPPs) enable ultrafast energy transport at metal surfaces on the nanoscale,[1–3] facilitating, for example, plasmonic nanocircuits,[4–7] plasmonic vortices,[8,9] and nanofocusing.[10–15] Conical tapers represent the quintessential structure for adiabatic nanofocusing,[10,11,16–19] harnessing the efficient coupling of propagating SPPs to a local apex excitation, which is accompanied by a slowing of the SPP group velocity.[10,11,19] The strong confinement of the plasmonic focus gives rise to various linear and nonlinear optical processes, with applications in microscopy and spectroscopy.[20–26]

Grating coupling was established for the excitation of SPPs on nanotapers,[13] providing a spatial separation of the far-field from near-field components and supporting background-free scanning probe techniques.[21,23,25,27–29] Moreover, the nanofocusing of SPPs by femtosecond laser pulses was shown to induce the emission of ultrashort electron pulses,[24–26] contributing to the development of compact ultrafast electron guns24,[30–33] and time-resolved point projection microscopy.[26,34] Such electron sources provide a complementary approach to localized electron emission driven by direct apex illumination.[35–42]

For each of these approaches, a detailed characterization of the underlying plasmon dynamics is desired to address questions regarding, e.g., the optimization of the excitation process,[43] the characterization of different plasmon modes,[44] and the reflection of SPPs near the apex.[45,46] SPP Propagation times and dispersion were recently investigated on thin nanorods[47–49] and plasmonic nanotapers[18] by interferometric means.

Photoelectron streaking is another powerful technique to access time-dependent electric fields. Prominently applied in attosecond science,[50,51] all-optical streaking was established for the analysis of electromagnetic transients, with further applications in the terahertz (THz) frequency range.[52–54] Streaking spectroscopy was proposed for the phase-resolved study of spatiotemporal near-field evolutions within nanostructures.[55–59] Experimentally, near-field streaking at nanostructures was first demonstrated at THz frequencies for metal nanotips[54,60,61] and recently implemented using attosecond pulses.[62,63] Besides the use in characterizing time-dependent near-fields, the strong field enhancement and local confinement in nanostructures have enabled the all-optical control and compression of photoelectron pulses.[54,61,64]

In this work, we apply THz near-field streaking to a plasmonic nanofocusing geometry and study the spatiotemporal SPP propagation on the shaft of a conical nanotip. THz streaking on plasmonic nanotapers combines the advantages of plasmon-induced photoemission, i.e., the spatial separation of excitation and emission site, with the capability to actively manipulate electron pulses. Comparing spectrograms employing SPP-mediated emission with direct apex excitation, our measurements yield the absolute propagation time from the grating to the apex.

The experimental scheme is depicted in Fig. 1(a), showing a scanning electron microscopy (SEM) image of the tip used in the experiment, overlaid with a sketch of the plasmon and THz excitation. The nanotips are produced by electrochemical etching of annealed gold wires,[21,23,24] and the grating coupler is milled into the tip shaft by a focused ion beam. Femtosecond near-infrared (NIR) pulses (duration, 50 fs; center wavelength, 800 nm; and repetition rate, 1 kHz) from an amplified Ti:sapphire laser system are used for the photoelectron emission from the nanostructure and the generation of single-cycle THz transients. The THz transients are generated with the AC bias method[65] by the major part of the NIR pulse energy (1.7 mJ combined energy in the fundamental and second harmonic of the NIR pulses) in a light-induced air-plasma, which allows for a control of the incident field strength and the carrier-envelope phase. Focused by a parabolic mirror of the focal length of 2.54 cm onto the tip apex, THz field strengths up to 100 kV/cm can be reached in this setup. The present measurements employ an incident field strength of 7 kV/cm.

A weak part of the NIR pulses (pulse energy, approx. 60 nJ) is focused onto a gold nanotip by a 15 cm lens to a focus of approx. 25 µm diameter, inducing photoelectron emission either by direct apex excitation or via nanofocused SPPs, created in a grating on the tip shaft [distance from the apex: 50 µm, cf. Figs. 1(b) and 1(c)]. The precise focus position on the tip was identified by spatial scans, as in Ref. 24. Most efficient SPP excitation was found for illumination near the grating edge. Thus, we do not observe any retardation of the plasmon velocity due to multiple scattering within or transmission through the grating.[66]

For both excitation conditions, the emitted electrons are streaked in a THz near-field, and the electron kinetic energy distributions as a function of relative THz-NIR pulse delay are acquired with a time-of-flight spectrometer. As discussed in our previous work, the resulting streaking spectrograms contain detailed information about the spatiotemporal electron dynamics in the near-field.[54,60,61]

The recorded spectrograms are displayed in Fig. 2. In the graphs, increasing delays correspond to the NIR pulse arriving later at the tip apex. The spectrogram in Fig. 2(a) is recorded with the NIR focus placed on the tip apex. To obtain the spectrogram of plasmon-induced photoemission [cf. Fig. 2(b)], the NIR beam is instead directed to the grating by tilting a mirror in about 40 cm distance from the focusing lens, holding the tip position constant. Both spectrograms exhibit the same overall shape and share even fine details (e.g., the enhanced feature at minimum energy and the following photocurrent suppression), which implies very similar conditions for the emission and propagation of the photoelectrons. The narrow spectral width in both direct and grating induced photoemission at negative delays implies that the emission process is multiphoton photoemission in the perturbative regime, consistent with estimates of the local field strengths. The spectrogram for grating excitation is shifted to negative delays with respect to that for apex excitation, which implies that the external NIR path length needs to be shortened in order to overlap the two pulses and access the same feature in the THz transient. Thus, the delay shift from Figs. 2(a) to 2(b) primarily represents the additional time required for the SPP wave packet to propagate from the grating to the apex.

We determine the delay shift between the spectrograms in Figs. 2(a) and 2(b) to $\Delta t_{streak}=-146$ fs by the maximum in their cross-correlation. The cross-correlation of the electron yield is computed separately for each energy, and the overall delay shift is found from an average of these values. To obtain the propagation time of the plasmon polaritons, we also need to take into account the change in optical path length Δl due to the repositioning of the NIR focus from the optical grating to the apex [see Fig. 2(c)]. The corresponding contribution $\Delta t_l=\Delta l/c_0=-27$ fs is measured by electro-optic sampling [cf. Fig. 2(d)], replacing the tip with a 500 µm thick ZnTe crystal. Hence, the pure SPP propagation time, more specifically the group delay time, amounts to $\Delta t_{SPP}=-(\Delta t_{streak}+\Delta t_l)=173$ fs. For the 50 µm distance between the optical grating and the tip apex, this corresponds to an average plasmon velocity of 96 % of the vacuum speed of light.

We simulated the propagation of SPPs along the tip shaft using an adiabatic model, which describes the plasmon velocity as a function of the taper diameter[11,19] (dielectric constants from Ref. 67). The results of the simulation are summarized in Fig. 3. The taper radii along the tip shaft [cf. Fig. 3(a)] are extracted from the SEM image in Fig. 1(a) to compute the local group velocity (solid blue) and propagation time (solid red) of the plasmons propagating from the grating to the tip apex [Fig. 3(b)]. In the last few micrometers close to the apex, the decreasing taper radius influences the dispersion relation, which leads to a deceleration of the SPP wave packet, but has only a marginal effect on the integrated group delay time. The simulated total propagation time from the grating to the tip apex amounts to 184 fs [Figs. 3(b) and 3(c)], which is in good agreement with our experimental results. The small difference most likely stems from deviations in the effective NIR beam path length not accounted for by the electrooptic sampling measurement, caused by, e.g., a minor relative tilt between the tip surface and the ZnTe crystal. The angular precision of the tip alignment was approximately 2°, corresponding to an error of 6 fs.

Figure 3(c) plots the dispersion of the plasmon group delay time, showing a decrease with the incident wavelength that corresponds to a group delay dispersion of 24 $fs^2$ (chirp of −0.0705 fs/nm) at a wavelength of 800 nm. In our experiment, the spectral density of the excitation pulse has a full-width at half-maximum (FWHM) of 26 nm. For a bandwidth-limited pulse, the increase in pulse duration due to the group delay dispersion would be less than 1 fs and is thus negligible.

In addition to the plasmon propagation, the streaking spectrograms give information on the influence of the plasmonic excitation on the electron distribution in the gold apex of the tip. The flat horizontal feature at higher energies and negative delays in Figs. 2(a) and 2(b) arises from THz-induced hot-electron tunneling.[60] The decay of this tunneling signal for negative delays signifies the evolution of the hot electron temperature in the apex region. The almost identical tunneling signal for both excitation conditions implies very similar electron dynamics for both types of excitation.

In conclusion, we employed THz streaking to measure the group delay time of SPPs upon nanofocusing. Generally, the experimental approach yields access to the relaxation of hot carrier distributions, SPP propagation times and dispersion, and the duration of photoelectron pulses, key features in the application of plasmonic nanotapers for time-resolved electron imaging. In the future, THz streaking can be used to study spatiotemporal field evolutions and carrier dynamics also in other structures, such as double-ended optical antennae or plasmonic nanoarrays, mapping SPP propagation with nanometer spatial and femtosecond temporal resolution.

We gratefully acknowledge funding by the Deutsche Forschungsgemeinschaft (DFG) (SFB-1073, Project C4) and by the European Research Council (ERC-StG. “ULEED,” Project ID: 639119). We thank Karin Ahlborn for help in tip fabrication.

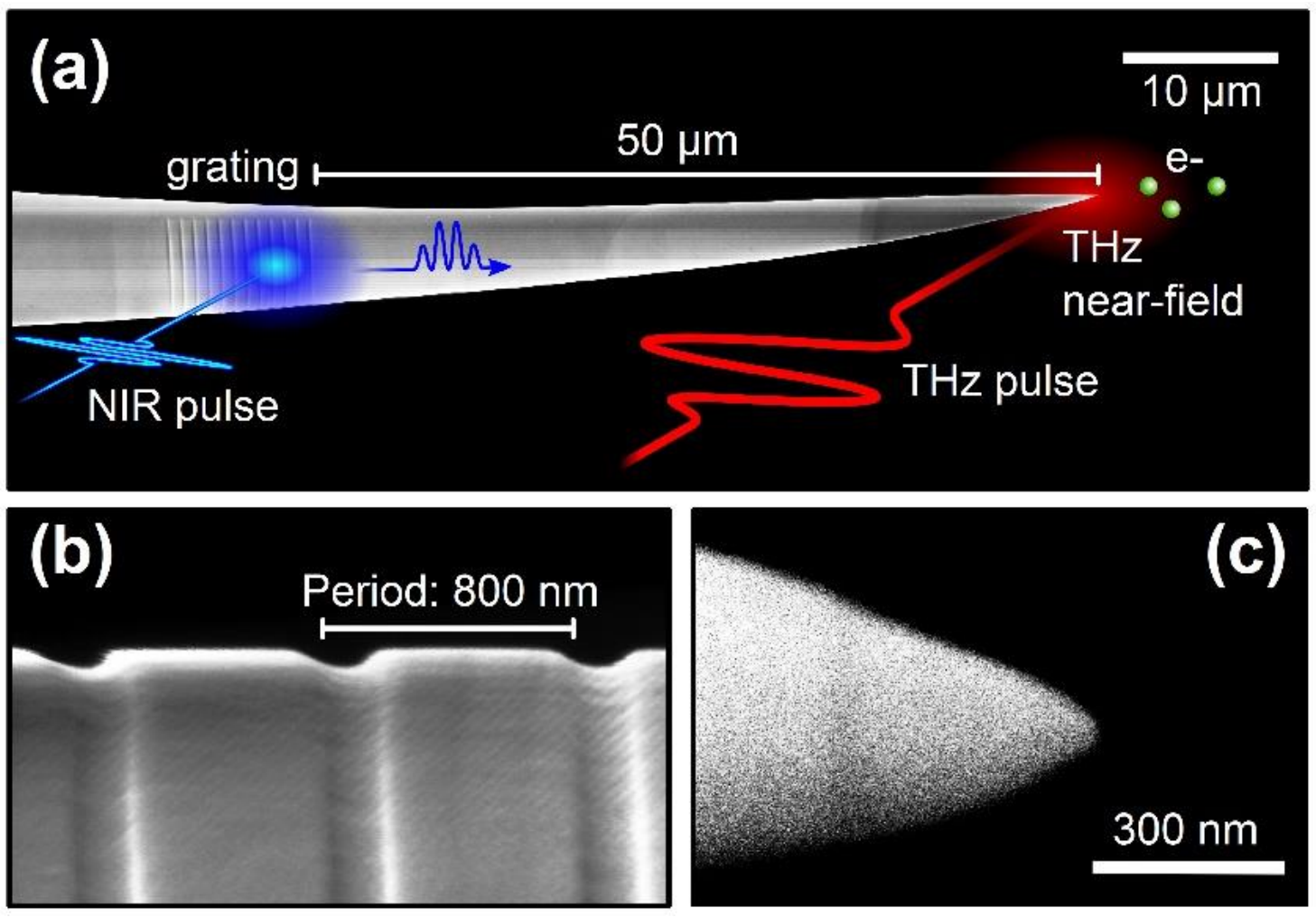

**FIG. 1.** Scanning electron microscopy images of the nanotip used in the experiment. (a) Micrograph of the nanotip including a sketch of the experiment. (b) Close-up of the grating structure that is optimized for incident light of 800 nm. (c) Close-up of the tip apex (radius of curvature: 20 nm).

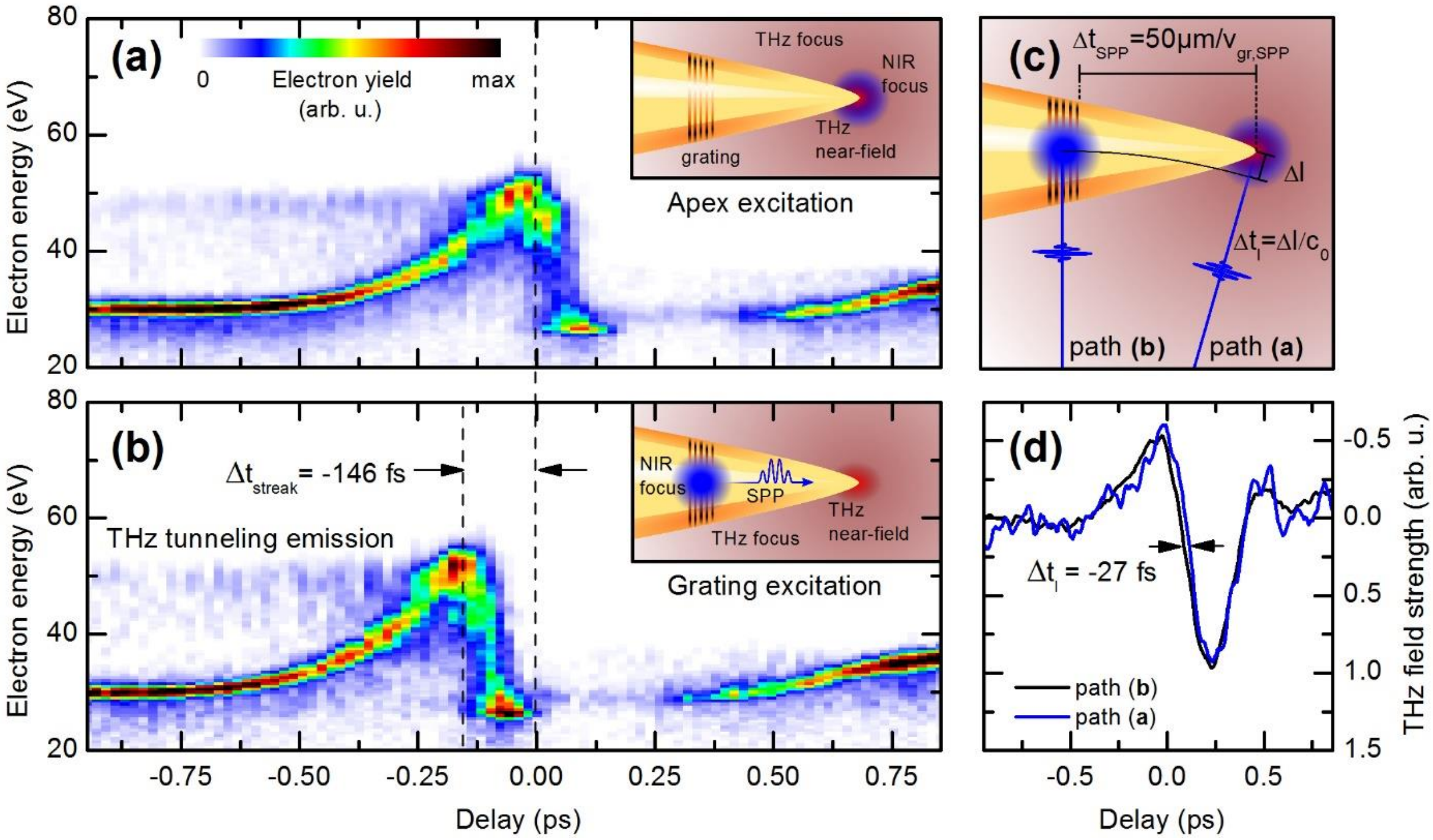

**FIG. 2.** Measuring the plasmon propagation time. (a) Streaking spectrogram with the NIR beam pointing to the tip apex. (b) Streaking spectrogram with plasmon excitation at the grating coupler (see sketch). In (a) and (b), the incident THz field strength was approx. 7 kV/cm, and a static bias voltage of $U_{bias}$ = -30 V was applied to the tip. (c) Sketch: change in the optical path length from configuration (a) to (b). (d) Electrooptic sampling traces for the configurations in (a) and (b).

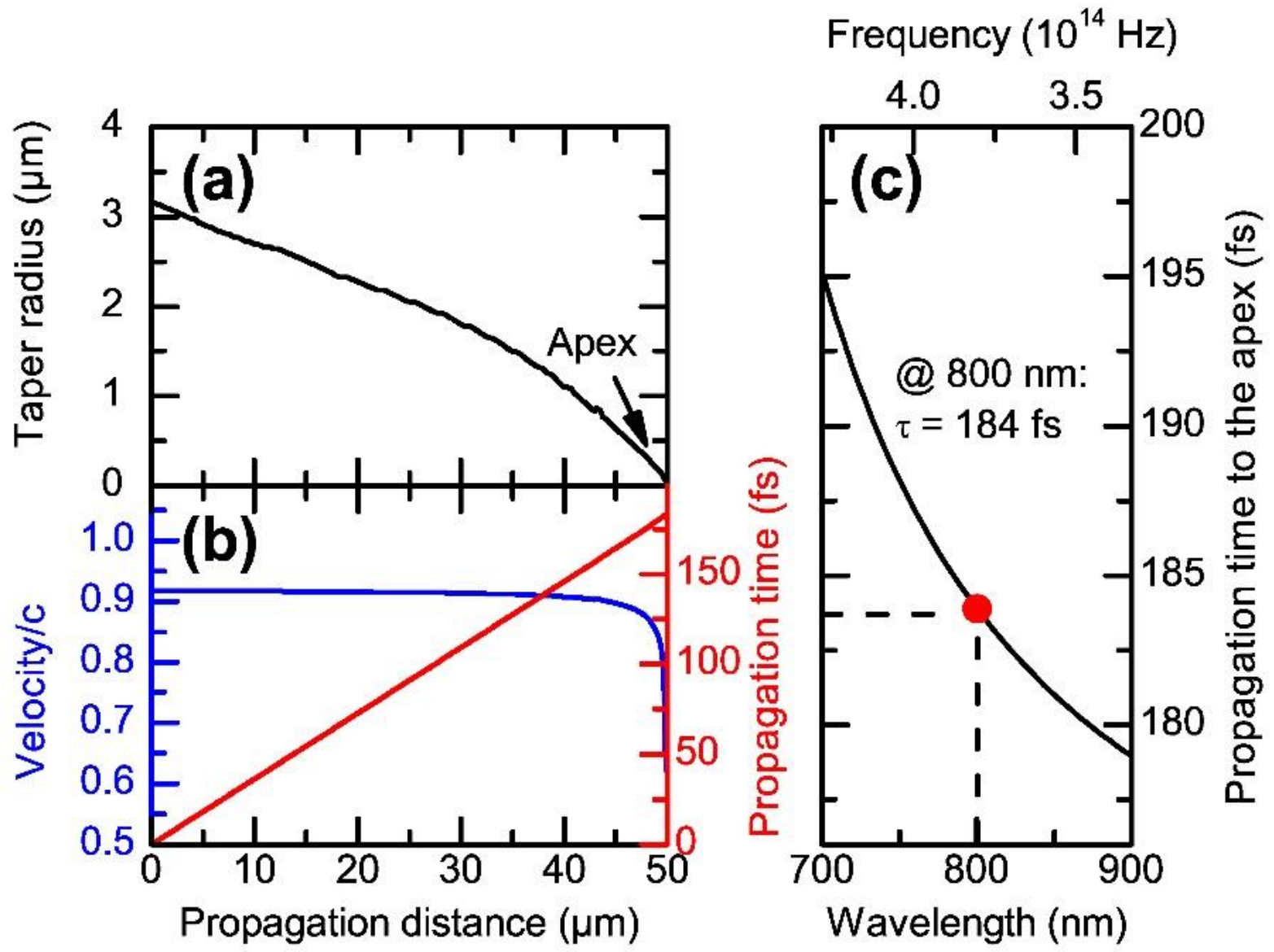

**FIG. 3.** Simulation results: (a) taper radius as a function of the propagation distance starting from the grating coupler (propagation distance, 0 µm) and ending at the apex (50 µm). The values are extracted from the SEM image shown in Fig. 1. (b) Plasmon velocity (blue, left scale) and propagation time (red, right scale) as a function of the propagation distance (incident/SPP wavelength: 800 nm/780 nm). (c) Propagation time from the grating to the apex for different excitation wavelengths.